\documentclass[pre,showpacs,showkeys,preprintnumbers,amsmath,amssymb,superscriptaddress,twocolumn,floatflix]{revtex4-1}
\usepackage[english]{babel}
\usepackage[dvipsnames]{xcolor}
\usepackage{makeidx} 
\usepackage{graphicx} 
\usepackage{dcolumn} 
\usepackage{array} 
\usepackage{amssymb} 
\usepackage{amsmath}
\usepackage{textcomp}
\usepackage{multirow}
\usepackage{subfigure}
\usepackage{eucal}
\usepackage{mathrsfs}
\usepackage{dsfont}
\usepackage{physics}
\usepackage{amsfonts} 
\usepackage{bm} 
\usepackage{float}
\usepackage{soul} 
\usepackage{ulem} 

\begin{document}

\title{Work and efficiency fluctuations in a quantum Otto cycle with idle levels}

\author{Maron F. Anka}\email{maronanka@id.uff.br}
\author{Thiago R. de Oliveira} \email{troliveira@id.uff.br}
\author{Daniel Jonathan}\email{djonathan@id.uff.br}

\affiliation{Instituto de F\'{i}sica Universidade Federal Fluminense - Av. Gal. Milton Tavares de Souza s/n, 24210-346 Niter\'{o}i, Rio de Janeiro, Brazil.}

\date{\today}
 
\begin{abstract}

We study the performance of a quantum Otto heat engine with two spins
coupled by a Heisenberg interaction, taking into account not only
the mean values of work and efficiency but also their fluctuations.
We first show that, for this system, the output work and its fluctuations are directly related to the magnetization and magnetic susceptibility of the system at equilibrium with either heat bath. We analyze the regions where the work extraction can be done with low relative fluctuation for a given range of temperatures, while still achieving an efficiency higher than that of a single spin system heat engine. In particular, we find that, due to the presence of `idle' levels, an increase in the inter-spin coupling can either increase or decrease fluctuations, depending on the other parameters. In all cases, however, we find that the relative fluctuations in work or efficiency remain large, implying that this microscopic engine is not very reliable as a source of work. 

\end{abstract}

\maketitle

\section{Introduction} \label{intro}
 
Quantum heat engines (QHEs) \cite{quan2007quantum,quan2009quantum} are rich scenarios to study the interplay between quantum physics and thermodynamics. Like their classical counterparts, they broadly consist of cyclical or steady-state processes in which heat is extracted from a high-temperature source, partially converted into useful work, and partially dumped into a lower-temperature sink. In a QHE, however, the `working substance' (WS) that interacts with the heat reservoirs is described using quantum mechanics, and can exhibit features such as discrete energy levels, energy-basis coherence, or entanglement between its constituent subsystems. A large amount of work has been devoted to studying the implications of these features for engine performance  \cite{zhang2007four,wang2012efficiency,elouard2018efficient,camati2019coherence,niedenzu2019concepts,chand2021finite,xiao2022thermodynamics,xiao2022finite}.

In particular, much attention has been recently directed toward understanding the quantum-originated fluctuations in these engines' work output and efficiency, which differ from those described using only classical stochastic thermodynamics. For example, it has been shown that a quantum Otto cycle can operate close to the Carnot efficiency without sacrificing power \cite{campisi2016power} when its WS is near a second-order phase transition.
In this regime, however, the work output presents large fluctuations. Similarly, Ref.~\cite{holubec2017work} studied a class of quantum Stirling cycles where the work fluctuations increase with the number $N$ of interacting subsystems, instead of decreasing with $1/\sqrt{N}$ as expected for large systems working away from the critical point of phase transitions. Later, Miller \textit{et. al.} \cite{miller2019work,scandi2020quantum} generalized an important result from classical stochastic thermodynamics to the quantum realm, finding a quantum correction to the work fluctuation-dissipation-relation that takes into account strictly quantum features.  

A particularly convenient type of QHE for studying such fluctuations is the quantum Otto cycle \cite{geva1992,kieu2004,quan2007quantum,camati2019coherence,pena2020otto,chand2021finite,xiao2022finite}, composed of four processes on a quantum WS: two isoentropic strokes, during which there is no heat exchange and the WS evolves unitarily, and two `isochoric' strokes, during which the WS exchanges heat either with a `hot' or with a `cold' external bath, but no work is performed. Important properties of quantum Otto cycles, such as work extraction processes, efficiency and power output, among others, have been extensively analyzed  \cite{wang2009thermal,thomas2011coupled,wu2014efficiency,zheng2014work,altintas2015general,ivanchenko2015quantum,wang2015efficiency,campisi2016power,ccakmak2017special,thomas2017implications,chand2017measurement,hewgill2018quantum,deffner2018efficiency,camati2019coherence,peterson2019experimental,oliveira2021efficiency,denzler2020efficiency,pena2020otto,denzler2021largedevs,denzler2021nonequilibrium,fei2022efficiency,xiao2022finite}. 

One point of contention in this analysis has been over the appropriate definition of the notion of cycle efficiency. In most studies of quantum heat engines, this quantity is taken, by extension of the classical definition, as the ratio of the \textit{mean values} of the total extracted work $W$ and of the heat $Q_h$ absorbed from the hot bath in a cycle \footnote{We adopt the convention that $\langle W \rangle < 0$ corresponds to work being extracted from the engine.}  
\begin{equation}
\eta_{th} \equiv - \dfrac{\langle W \rangle}{\langle Q_h \rangle}.\label{themodeff}
\end{equation}
We will refer to this quantity as the \textit{thermodynamic efficiency}.
However, in a quantum engine, both $W$ and  $Q_h$ are stochastic quantities and may present fluctuations that are by no means negligible \cite{campisi2014fluctuation}. The same is therefore true of the ratio
\begin{equation} \label{eq:stochasticeff1}
\eta_s \equiv - \dfrac{W}{Q_{h}},
\end{equation}
which we henceforth refer to as the \textit{stochastic efficiency}.

In general, the average value $\langle \eta_s \rangle$ does not coincide with $\eta_{th}$. Even worse, $\eta_s$ diverges whenever $Q_h = 0$ and $W \neq 0$, something that cannot be discarded if $W$ and $Q_h$ are stochastic variables. Indeed, as Denzler \textit{et. al.} shown \cite{denzler2020efficiency}, this happens with finite probability already in the simplest case of an Otto cycle with a WS consisting of a single qubit, for any finite-time compression/expansion strokes. In this case, $\eta_s$ does not even have a well-defined average or higher statistical moments. Nevertheless, they also found that in the adiabatic limit of very slow strokes, the probability of divergence goes to zero, and in fact, $\langle \eta_s \rangle$ becomes deterministic and coincides with $\eta_{th}$ \cite{denzler2020efficiency}. 

Outside this limit, reasonable values for the moments of $\eta_s$  may arguably still be obtained by ad-hoc truncation of the range of values over which averages are performed. For example, this was done in Ref. \cite{denzler2021nonequilibrium}, which analyzed a quantum Otto engine realized experimentally in a nuclear magnetic resonance setup. This work also studied the connection between work-heat correlations and the efficiency and entropy production statistics of the engine. 
The stochastic efficiency also seems to have reasonable large-deviation statistics after a very large number of cycles. Such an analysis was performed in Ref. \cite{denzler2021largedevs}, where among other things it was found that the relevant large deviation function for $\eta_s$ displays `universal' features  \cite{verley2014unlikely,verley2014universal,manikandan2019efficiency,vroylandt2020efficiency}, such as a trough at the thermodynamic efficiency and a peak at the Carnot efficiency. Once again, the adiabatic limit was found to be an exceptional case.

Nevertheless, it remains the case that $\eta_s$  generally has divergent moments. For this reason, Fei \textit{et. al.} \cite{fei2022efficiency} have introduced a different notion of efficiency, named the `scaled fluctuating efficiency', defined as the ratio of the stochastic work and the \textit{mean} absorbed heat
\begin{equation}
\eta_{s'} = - \dfrac{W}{\langle Q_h \rangle}.
\label{stochasticefficiency2}
\end{equation}

Unlike the stochastic efficiency, this quantity is non-divergent, since, by the Second Law, $\langle Q_h \rangle > 0$ necessarily for heat engine cycles. Moreover, its mean value is equal by construction to  $\eta_{th}$. We can therefore use $\eta_{s'}$ as a stochastic version of $\eta_{th}$. In this paper, for simplicity of language, whenever there is no risk of confusion we will refer to $\eta_{s'}$ itself as `the thermodynamic efficiency'. However, it is important to note that all the fluctuations in $\eta_{s'}$  are inherited only from those of the work variable $W$. In particular, both quantities share the same  \textit{relative} fluctuations, or `coefficient of variation' \cite{upton2014dictionary}:
\begin{equation}\label{varcoefficient}
\frac{\sigma_{\eta_{s'}}^2}{\eta_{th}^2 } = \frac{\sigma_{W}^2}{\langle W \rangle^2}
\end{equation}
where $\sigma$ refers to the respective variances (see eq.(\ref{mean})). Using this definition, Fei \textit{et. al.} \cite{fei2022efficiency} studied the stochastic nature of the quantum Otto engine running with a working substance composed of a single harmonic oscillator. In particular, they were able to verify that these relative fluctuations satisfy thermodynamic uncertainty relations (TURs) \cite{barato2015thermodynamic,timpanaro2019thermodynamic,proesmans2017discrete,horowitz2020thermodynamic}.

In this paper, we explore the fluctuations of a quantum Otto cycle based on a pair of interacting spins in the presence of an external magnetic field \cite{thomas2011coupled, oliveira2021efficiency,anka2021measurement,cherubim2022nonadiabatic}.  Unlike the model studied by Fei \textit{et. al.}, in this case, the spin-spin coupling causes the appearance of `idle' energy levels, i.e., energy levels that do not shift during the cycle. This feature has a significant effect on the engine's thermodynamical properties \cite{oliveira2021efficiency}.  In particular, the coupled-spin Otto engine may function with higher thermodynamic efficiency $\eta_{th}$ when compared to classical Otto engines with an equivalent compression ratio \cite{thomas2011coupled, oliveira2021efficiency}, or also to quantum Otto engines based on uncoupled spins, or harmonic oscillators \cite{thomas2011coupled,kosloff2017quantum}. Here we seek to analyze the effect of the coupling also on the fluctuations of the cycle's work output and efficiency.
It is particularly interesting to ask whether the apparent advantage in the \textit{mean} efficiency survives when fluctuations are taken into account. In other words, we would like to understand whether, at least for some regions of parameter space, the coupled-spin Otto engine with idle levels can function above the classical Otto efficiency, while simultaneously displaying relatively low fluctuations. This is the main goal of this work.

The paper is organized as follows: we begin in section \ref{ottocycle} with a brief general review of quantum Otto cycles and the two-point measurement protocol (TPM). In section \ref{sec:Otto} we review the properties of the Otto cycle on our model system and also point out the relation between the physical properties of the system such as its magnetization and magnetic susceptibility with the cycle's work output and its fluctuations. In section \ref{workfluctuation} we explore how quantities of interest such as 
the mean output work, its fluctuation, and its relative fluctuations behave with respect to changes in the temperatures of the cold and hot reservoirs. In section \ref{eff} we explore the same issues for the thermodynamic efficiency ($\eta_{s'}$) and its fluctuation. We furthermore study the stochastic consequences of the TPM protocol for the efficiency's probability distribution. We also verify that the relative fluctuations satisfy a lower bound given by a thermodynamic uncertainty relation (TUR). In section \ref{conclusion} we give our final remarks and conclusions. 

\section{Otto cycle and the TPM protocol} \label{ottocycle}

Let us begin by reviewing the basic description of a quantum Otto cycle \cite{pena2020otto}. Consider a generic quantum WS, governed by some time-dependent 
Hamiltonian $H(\lambda(t))$, where $ \lambda(t)$ is an externally tunable parameter. The four strokes of a quantum Otto cycle can be briefly described as follows. First stroke: starting from a thermal equilibrium state at inverse temperature $\beta_c=~1/T_c$ ($k_B=1$), the WS evolves unitarily while the external parameter is changed from $\lambda_i \rightarrow \lambda_f$. During this process, the system is isolated from the thermal reservoirs, so the only contribution to its energy change is in the form of work $W_1$ \cite{alicki1979quantum,quan2007quantum}. In this paper, we assume that this is an ideal quantum adiabatic evolution, where there is no change in the initial energy occupation probabilities $p_n^c$, while the energy eigenvalues and eigenstates evolve smoothly from those of $H^i = H(\lambda_{i})$ to those of $H^f = H(\lambda_{f})$. This is merely a matter of simplicity - it would be possible to drop this assumption and analyze finite-time (nonadiabatic) unitary strokes (see, e.g. \cite{solfanelli2020nonadiabatic, cherubim2022nonadiabatic} and references therein). Second stroke: The WS thermalizes with a `hot' heat bath at inverse temperature $\beta_h = 1/T_h < \beta_c$. No work is done in this step, since $\lambda$ is fixed at $\lambda_f$, and only heat $Q_h$ is exchanged by the system with the bath. Third stroke: This process is similar to the first stroke. Here, the external parameter is changed back to the initial value, $\lambda_f \rightarrow \lambda_i$, and the occupation probabilities remain fixed at $p_n^h$. Only work $W_2$ is performed and no heat is exchanged. Fourth Stroke: This is another isochoric process in which the WS is put in thermal contact with a `cold' heat bath at inverse temperature $\beta_c$, exchanging energy $Q_c$. Once again, no work is produced. After thermalization, the system returns to its initial state, closing the cycle.

Before we can analyze the statistics of the work distribution in this cycle, we must first pick a definition for this quantity. Despite numerous efforts \cite{talkner2007fluctuation,roncaglia2014work,allahverdyan2014nonequilibrium,alonso2016thermodynamics,deffner2016quantum,baumer2018fluctuating,micadei2020quantum}, no unique, universally accepted quantum-mechanical definition of work yet exists. However, the divergence of definitions applies mainly in situations where energy-basis coherence is involved. Since, in this paper, we only consider situations with no energy-basis coherence, we choose for convenience to use the two-projective-measurement (TPM) notion of work \cite{tasaki2000jarzynski,kurchan2000quantum,talkner2007fluctuation}. In this case one considers projective measurements in the energy basis of the system's Hamiltonian at the beginning and at the end of each adiabatic stroke. The difference between the two measured values is associated with the energy change within the quantum system and also (by energy conservation) with the work absorbed during that stroke. As already mentioned, this is a stochastic quantity, dependent on the result of both random measurements. 

For the first stroke (adiabatic stroke starting from equilibrium with the cold bath), the probability density of extracting an amount $W_1$ of work from the engine is then given by

\begin{equation}
\label{eq:PW1}
P(W_1) = \sum_{n,m} \delta[W_1 - (E^f_m - E^i_n)] p_{n\rightarrow m} p^i_n(\beta_c).
\end{equation}
Here $p_{n\rightarrow m} \equiv |\bra{m}U_{exp}\ket{n}|^2$ is the transition probability between the initial and final energy eigenstates after the unitary evolution $U_{exp}$,  $E^{\alpha} \equiv E(\lambda_{\alpha})$, with $\alpha \in \{{i,f}\}$, are the corresponding initial and final energy eigenvalues, $p^i_n(\beta_c) = e^{-\beta_c E^i_n}/Z^i$ is the initial occupation probability with the partition function $Z^i$ at thermal equilibrium with the cold heat bath, and $\delta(x)$ is the Dirac delta function  \footnote{Note that because the working substance is a finite-dimensional quantum system, the range of possible values of $W_1$ is in fact finite. Nevertheless, it is convenient to represent it as a (singular) probability density over a continuum of values. Equivalently, one can also interpret eq.(\ref{eq:PW1}) as a discrete probability distribution, by reinterpreting the symbol $\delta[x]$ as the discrete delta function ($=1$ if $x =0$ and $=0$ otherwise)}.

Similarly, we can write the joint probability density for $Q_h$ and $W_1$, where $Q_h$ is the heat exchanged by the system while thermalizing with the hot thermal bath during the second stroke (the hot isochoric).

\begin{equation}
\label{eq:PQh}
P(W_1, Q_h) = \sum_{l,m,n} \delta[Q_h - (E^f_l - E^f_m)] p_{m\rightarrow l} \;p_{nm}(W_1).
\end{equation}
Here $p_{m\rightarrow l}$ is the transition probability between energy eigenstates during the (non-unitary) thermalization process, and $p_{nm}(W_1) \equiv \delta[W_1 - (E^f_m - E^i_n)] p_{n\rightarrow m} p^i_n(\beta_c)$ is the summand in eq.(\ref{eq:PW1}) above.

In the third stroke, work $W_2$ is performed to bring the parameter $\lambda$  back to its initial value. The joint probability density for $W_1, Q_h$ and $W_2$ is

\begin{equation} \label{eq:PW1QhW2}
P(W_1,Q_h,W_2) = \sum_{j,l,m,n} \delta[W_2 - (E^i_j - E^f_l)] p_{l\rightarrow j} \; p_{lmn}(W_1,Q_h).
\end{equation} 
Here $p_{l\rightarrow j} = |\bra{j}U_{com}\ket{l}|^2$ is the transition probability described by the compression unitary evolution $U_{com}$, and 
$p_{lmn}(W_1,Q_h) \equiv \delta[Q_h - (E^f_l - E^f_m)] p_{m\rightarrow l} \;p_{nm}(W_1)$ is the summand in eq.(\ref{eq:PQh}) above. The marginal probability density for $W_2$, which can be obtained by integrating $P(W_1,Q_h,W_2)$, is analogous to eq.(\ref{eq:PW1}):
\begin{equation}
\label{eq:PW2}
P(W_2) = \sum_{j,l} \delta[W_2 - (E^i_j - E^f_l)] p_{l\rightarrow j} p^f_l,
\end{equation}
where $ p^f_l$ is the probability of the system being found with energy $E_l^f$ at the end of stroke 2. 

The probability density for the `scaled fluctuating efficiency' $\eta_{s'}$ is then
\begin{align}
P(\eta_{s'}) &= \int dW_1 dQ_h dW_2 P(W_1,Q_h,W_2) \delta\left(\eta_{s'} + \frac{W_1 + W_2}{\langle Q_h \rangle}\right).
\label{stochasticefficiency}
\end{align}
Note that, since $\eta_{s'}$ depends only on the average absorbed heat $\left\langle Q_h \right\rangle$, then the integration in $Q_h$ is trivial, and this density can be written only in terms of the 
the marginal joint density $P(W_1, W_2) = \int dQ_h P(W_1,Q_h,W_2)$:
\begin{align}
P(\eta_{s'}) &= \int dW_1 dW_2 P(W_1,W_2) \delta\left(\eta_{s'} + \frac{W_1 + W_2}{\langle Q_h \rangle}\right).
\label{stochasticefficiency2}
\end{align}

Note also that, in the limit where the system reaches full thermal equilibrium with the hot heat bath at the end of stroke 2, then $p^f_l = p^f_l(\beta_h) = e^{-\beta_h E^f_l}/Z^f$ in eq.(\ref{eq:PW2}). In addition,  in eq.(\ref{eq:PQh}),  $p_{m\rightarrow l}~=~p^f_l(\beta_h)$ becomes independent of $m$. Physically, this means that the system loses all memory of the previous strokes, and  $P(W_1, W_2)$ factorizes into $P(W_1) P(W_2)$. For simplicity, in what follows we will assume we are in this regime.

Finally, one can evaluate the $n^{th}$ moments and the variance of any stochastic quantity $X$, with probability density $P(X)$, in the usual way

\begin{equation}
\begin{array}{cc}
\langle X^n \rangle &= \int dX X^n P(X),  \\
\sigma^2_X =& \langle X^2 \rangle - \langle X \rangle^2.
\end{array}
\label{mean}
\end{equation}

\begin{figure*}[t]
\centering
\includegraphics[width=0.9\textwidth]{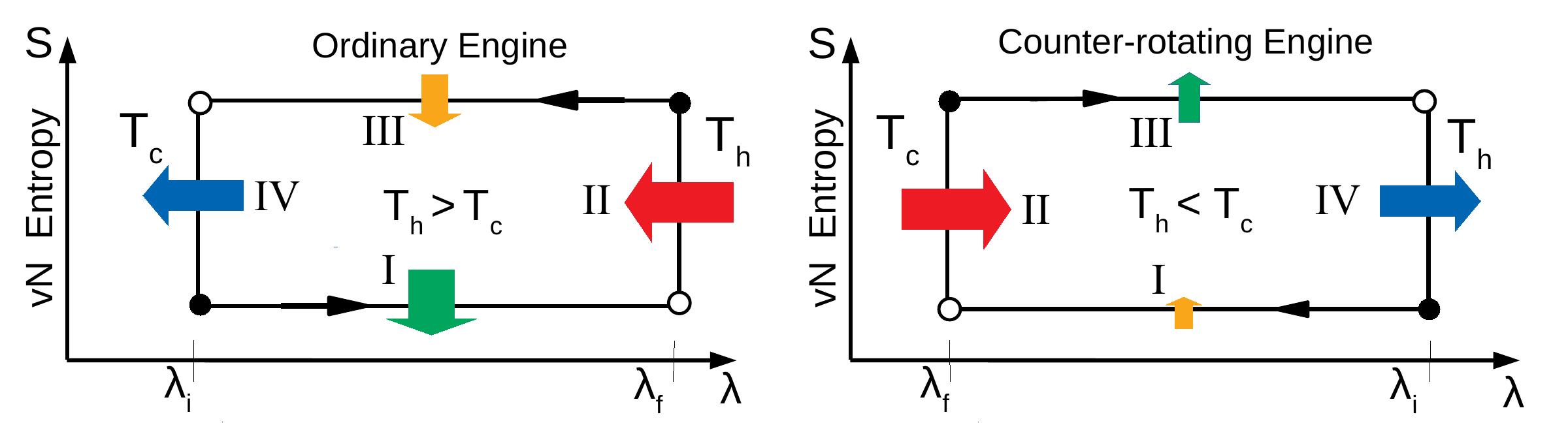}
\caption{(Color online) Quantum Otto engine cycles. $S$ is the von Neumann entropy, while $\lambda$ is some external parameter of the Hamiltonian, e.g. an external field intensity, whose increase widens the system's energy gaps. Vertical (isochoric) strokes involve only heat exchange, represented by horizontal colored arrows. Horizontal (adiabatic) strokes involve only work exchange, represented by vertical colored arrows. Filled circles represent thermal equilibrium states.
(a) An `ordinary' quantum Otto engine  \cite{geva1992,kieu2004,quan2007quantum,oliveira2021efficiency} operates in an anti-clockwise cycle in this representation. (b) If the working substance contains `idle' levels, unaffected by the change in $\lambda$, an Otto engine may also operate clockwise \cite{oliveira2021efficiency}. In this case, the two baths exchange their roles, i.e., '$T_c$' becomes the temperature of the hot bath. Other modes of operation (e.g., refrigerators) are also possible in either cycle sense, see \cite{oliveira2021efficiency}. \color{black}}
\label{Fig:Otto}
\end{figure*}

\section{Otto cycle with idle levels}\label{sec:Otto}

In this section, we review some properties
of a quantum Otto engine whose working substance has two kinds of levels: `working' levels, which shift proportionately to the external parameter $\lambda$, and `ìdle' levels, which remain fixed \cite{oliveira2021efficiency}. The population in these levels do not therefore contribute to work exchanges during the unitary strokes. As explored in detail in Ref. \cite{oliveira2021efficiency}, this leads to Otto cycles with several counter-intuitive properties, including:
\begin{itemize}
    \item The ability to function as a heat engine in two distinct modes (Fig.~\ref{Fig:Otto}): in the `ordinary' engine mode, the first and third strokes (as described in Section \ref{ottocycle}) are respectively expansion and compression strokes: $\lambda_f > \lambda_i$, $\left\langle W_1 \right\rangle<0$ and $\left\langle W_2 \right\rangle>0$. In the `counter-rotating' engine mode, the opposite occurs: $\lambda_f < \lambda_i$, $\left\langle W_1 \right\rangle>0$ and $\left\langle W_2 \right\rangle<0$. Note that in both cases there is a net extraction of work ($\left\langle W \right\rangle<0$), even though in each case the cycle is performed in a different sense.
    \item A thermodynamic efficiency $\eta_{th}$ that can reach values higher than the standard macroscopic Otto expression: $\eta_{th} > \eta_{Otto} \equiv 1 - \lambda_{\min} / \lambda_{\max}$.
   \item Anomalous dependence of the thermodynamic efficiency on the temperature difference between the heat baths. In other words it is possible for $\eta_{th}$ to increase as $T_h - T_c$ decreases, and vice-versa.  
\end{itemize}
These properties do not depend on the assumptions of perfect adiabatic strokes or exact thermalization \cite{oliveira2021efficiency,cherubim2022nonadiabatic}.

In this paper we are interested in investigating, in the same scenario, the fluctuations in both the work and the efficiency, as measured by $\eta_{s'}$. We are particularly interested in looking at how the coefficient of variation in eq.(\ref{varcoefficient}) depends on the cycle parameters. In particular, it is interesting to investigate under what conditions the cycle can function with high average efficiency while still maintaining small relative fluctuations.

\subsection{Model}

Following Ref.\cite{oliveira2021efficiency}, we choose as our test system a simple model of two interacting spin-1/2 qubits placed in a uniform, classically described magnetic field $h$ \cite{oliveira2021efficiency,cherubim2022nonadiabatic}. This field is assumed to be externally controllable and plays the role of the external parameter $\lambda$. We furthermore assume an isotropic Heisenberg interaction, so the system Hamiltonian reads

\begin{equation}
H = J (\Vec{S}_1 \cdot \Vec{S}_2) + h (S_{1z} + S_{2z}) -J/4~\mathds{1}.
\label{hamiltonian}
\end{equation}
Here $h \hat{z} $ is the magnetic field, $\Vec{S}_i$ is the spin operator of the i-th spin and $J$ is the coupling interaction between both qubits, with $J < 0$ and $J > 0$ being the ferromagnetic and antiferromagnetic regimes, respectively \footnote{We assume units such that Planck's constant and the magnetic moment of the spins are both numerically equal to 1 so that both $J$ and $h$ can be considered as measured in units of energy. Note that, as mentioned above $h$ here is the magnetic field, and not Planck's constant.}. The last term is an energy shift added so that one of the system's eigenenergies is zero. The full set of eigenergies and eigenstates are listed in Table \ref{heisenbergeigen} below.

\begin{table}[H]
\centering
\begin{tabular}{ |c|c|c|c|c| } 
\hline
Eigenvalues & Eigenstates  \\ 
\hline
$+h$ & $\ket{11}$  \\
\hline
$0$ & $(\ket{01} + \ket{10})/ \sqrt{2}$ \\
\hline
$-J$ & $(\ket{01} - \ket{10})/ \sqrt{2}$ \\
\hline
$-h$ & $\ket{00}$ \\
\hline
\end{tabular}
\caption{The four eigenvalues of the Hamiltonian $H$ with their associated eigenvectors.}
\label{heisenbergeigen}
\end{table}

Note that the energy level $E = -J$ is `idle', in the sense defined above. In addition, the eigenbasis is also insensitive to $h$. This implies that, for this particular model, the adiabatic strokes can, in fact, be realized in finite time with no loss of efficiency (i.e., there is no `quantum friction' \cite{oliveira2021efficiency}).

The work output and efficiency of quantum cycles based on this interaction have been studied by several authors, e.g.:  \cite{thomas2011coupled,altintas2015general,hewgill2018quantum,oliveira2021efficiency,cherubim2022nonadiabatic}. In particular, some of the present authors have studied this model for the cases of adiabatic and non-adiabatic processes \cite{oliveira2021efficiency,cherubim2022nonadiabatic}. These works analyzed the effects of the spin-spin coupling on the efficiency and on the regime of operation of the Otto cycle as a function of the bath temperatures (i.e., whether it performs the roles of heat engine, refrigerator, heater or accelerator \cite{solfanelli2020nonadiabatic}). In this paper, we focus only on the engine regime. Finally, the two different heat engine modes mentioned above (`counter-rotating' and `ordinary') are associated with whether or not the idle $E = -J$ energy level is the ground state  (i.e, whether or not $J > h$)  \cite{oliveira2021efficiency}. The ferromagnetic case, where $J<0$ also results in `ordinary' behavior, is qualitatively similar to the low-coupling regime of the antiferromagnetic case ($0 \leq J < h$), and so we will not discuss it here.

\subsection{Work and Magnetization} \label{sec:mag}

Before studying the engine cycle, let us first discuss some physical aspects of work exchanges in this system (see again Ref. \cite{oliveira2021efficiency} for further details). 

As mentioned above, in this cycle work is only exchanged during the unitary strokes \cite{alicki1979quantum,quan2007quantum}. In the first stroke, starting from thermal equilibrium at temperature $T_c$, the average work done on the system corresponds to the change in the average energy of the spins as the external field is changed from $h_i$ to $h_f$. Using the probabilities in eq.(\ref{eq:PW1}) and the energies in Table~\ref{heisenbergeigen}, one obtains

\begin{equation}
\langle W_1\rangle = - 2 \Delta h \frac{\sinh(x_c)}{Z_c},
\label{Eq DU}
\end{equation}
where $x_c = \beta_c h_i$, $Z_{c} = 1 + e^{\beta_{c}J} + 2 \cosh(x_c)$ is the partition function in thermal equilibrium with the cold bath at field $h_{i}$, and $\Delta h = h_f - h_i $. An analogous expression applies in the third stroke, but with initial field $h_f$ and temperature $T_h$,  $x_h = \beta_h h_f$, and $Z_{h}= 1 + e^{\beta_{h}J} + 2 \cosh(x_h)$. 
The total average work in the cycle is thus \cite{thomas2011coupled}

\begin{equation}
\langle W \rangle = \langle W_1\rangle + \langle W_2\rangle =2 \Delta h \Big[ \dfrac{1}{Z_{h}} \sinh(x_{h}) - \dfrac{1}{Z_{c}} \sinh(x_{c}) \Big].
\label{work}
\end{equation}

Note that work is extracted from the system as the external magnetic field is \textit{increased}, and vice-versa (in analogy to work being extracted when a gas is allowed to expand against a piston, and vice-versa). This happens because, although the magnetic-dependent energy levels shift equally in opposite senses, the lower level has a higher population. Note also that, for coupled spins, these quantum adiabatic processes in general drive the system out of thermal equilibrium \cite{quan2007quantum,oliveira2021efficiency}.

Let us now consider the fluctuations in the work output. Using again eqs.(\ref{eq:PW1}) and (\ref{eq:PW2}), and also eq.(\ref{mean}), one obtains that, in each unitary stroke:
\begin{equation}
\sigma^2_{W_i} = (\Delta h)^2 \left(
\dfrac{2 \cosh(x_{\alpha})}{Z_{\alpha}} - \dfrac{4 \sinh^2(x_{\alpha})}{Z^2_{\alpha}}\right),
\label{eq:fluc1}
\end{equation}
where $\alpha \in \{c,h\}$ refers to the temperature and magnetic field strength at the beginning of the stroke.

For the full cycle, the fluctuations in the total work $W = W_1 + W_2$ are, by definition, $\sigma^2_W = \sigma^2_{W_1} + \sigma^2_{W_2} + 2 ~Cov(W_1,W_2)$. However, the covariance $Cov(W_1,W_2)$  vanishes due to the perfect thermalization during the isochoric processes. Therefore:
\begin{equation}
\sigma^2_{W} = (\Delta h)^2 \sum_{\alpha \in \{c,h\}} 
\dfrac{2 \cosh(x_{\alpha})}{Z_{\alpha}} - \dfrac{4 \sinh^2(x_{\alpha})}{Z^2_{\alpha}}.
\label{eq:fluc}
\end{equation}

It must be emphasized that, despite the quantum nature of the working substance, at no point during the cycle is any coherence created between energy eigenstates. This means that all energy fluctuations are entirely thermal in origin (i.e., not due to quantum indeterminacy). In particular, in the limit where $x_\alpha \rightarrow \infty$ (i.e., $T_\alpha\rightarrow 0$), eq. (\ref{eq:fluc}) vanishes. 

Let us now discuss a physical interpretation for these fluctuations. Previous studies of work fluctuations in quantum Otto cycles \cite{campisi2016power,holubec2017work,denzler2021power} concluded that they could be associated with the heat capacity of the system, $C = \partial \langle H \rangle/\partial T$, calculated at thermal equilibrium. This occurred because, in those studies, \textit{all} energy levels of the working substance shifted during the adiabatic strokes. In contrast, for the working substance described by eq.(\ref{hamiltonian}), only some energy levels (those that depend on the magnetic field intensity $h$) shift during the adiabatic strokes. As a consequence, here the (average)  work done in these strokes may be associated with the  (average) magnetization  $\left. \langle M\rangle = -\partial F/\partial h\right|_T$, where $F = -T \ln Z$ is the free energy calculated at thermal equilibrium. Furthermore, the work fluctuations are associated not with the heat capacity, but with the magnetic susceptibility $\chi = \left.\partial \langle M\rangle/\partial h\right|_T$. 
To see this in more detail: note first that, for thermal states,
\begin{equation}
\langle M\rangle = - \sum_n p_n \dfrac{\partial E_n}{\partial h},
\label{mag}
\end{equation}
\begin{equation}
\chi = \beta \Big[ \sum_n p_n \Big(\dfrac{\partial E_n}{\partial h}\Big)^2 - \Big( \sum_n p_n \dfrac{\partial E_n}{\partial h}\Big)^2 \Big] - \sum_n p_n \dfrac{\partial^2 E_n}{\partial h^2}.
\label{susc}
\end{equation}
eq.(\ref{mag}) shows that  $\langle M\rangle$ corresponds to the expectation value of the magnetization observable $-\partial H / \partial h$. Meanwhile, in eq.(\ref{susc}), the first term is proportional to the variance $\sigma^2_M$ of this observable. The second term is the average value of the second derivative observable  $\partial^2 H / \partial h^2$, which vanishes for the  Hamiltonian $H$ in eq.(\ref{hamiltonian}). Thus, the susceptibility measures the fluctuations in the magnetization observable. Note that, more generally, the second term in eq.(\ref{susc}) is also negligible in the limit of small magnetic fields $( h \ll \beta^{-1} )$ \cite{reis2013fundamentals}.

Taking into account the energies in Table~\ref{heisenbergeigen}, and comparing eqs.(\ref{Eq DU}) and (\ref{mag}), we can now see that the  work extracted in each unitary stroke is proportional to the magnetization of the thermal equilibrium state immediately preceding that stroke. The total work is simply  
\begin{equation}
\langle W \rangle = \Delta h  \left(\langle M_h^f \rangle - \langle M_c^i \rangle\right) = \Delta h \Delta M,
\label{eq:workmag}
\end{equation}
where $\langle M_{c(h)}^{i(f)}\rangle$
are the average magnetizations of the system in thermal equilibrium with the cold (hot) bath at temperature $T_{c(h)}$, and external field $h_{i(f)}$, respectively, and $\Delta M \equiv \langle M_h^f \rangle - \langle M_c^i \rangle.$

Similar relations can be derived between the work fluctuations and the magnetic susceptibility: 
\begin{equation}
\sigma_{W_{1(2)}} = \Delta h\sigma_{M_{c(h)}^{i(f)}} = \Delta h\sqrt{T_{c(h)}\chi_{c(h)}}\;,
\end{equation} 
where $\chi_{c(h)}$ 
are the magnetic susceptibilities of the system in thermal equilibrium with the cold (hot) bath at temperature $T_{c(h)}$, respectively.

The relative fluctuations in each stroke are thus 
\begin{equation}
\dfrac{\sigma_{W_1}}{\langle W_1 \rangle} = -\dfrac{\sqrt{T_{c}\chi_{c}}}{\langle M_c^i \rangle}, \ \ \ \dfrac{\sigma_{W_2}}{\langle W_2 \rangle} = \dfrac{\sqrt{T_{h}\chi_{h}}}{\langle M_h^f \rangle}.
\end{equation}
For the full cycle, using also eq.(\ref{eq:fluc})
\begin{equation}
\dfrac{\sigma_W}{\langle W \rangle} = \dfrac{\sqrt{T_c \chi_c + T_h \chi_h}}
{\Delta M}.
\label{workfluct}
\end{equation}

In conclusion, eqs.(\ref{eq:workmag}) and (\ref{workfluct}) tell us that the total work produced by the coupled-spin system in a full Otto cycle, and also its fluctuations, can be inferred from measurements of its temperature, magnetization and magnetic susceptibility at thermal equilibrium with each bath, quantities that in principle should be experimentally accessible. We emphasize once again that this is true despite the fact that the system does not remain in thermal equilibrium during the operation of the cycle. 

\section{Work and its fluctuation}
\label{workfluctuation}

In this section, we explore how the work output of the engine, and also its fluctuations, vary depending on the temperatures of the external baths and on the parameters of the WS, in particular on the coupling strength $J$. We are mainly interested in finding parameter regions where the work output is large, but where relative fluctuations in this output are also small.

\subsection{Average Work}

\begin{figure}[t]
\centering
\subfigure{
    \includegraphics[width=4.1cm,height=4.1cm]{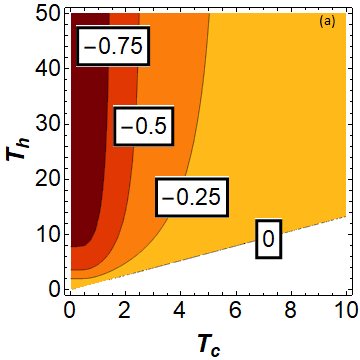}%
}
\subfigure{
\includegraphics[width=4.1cm,height=4.1cm]{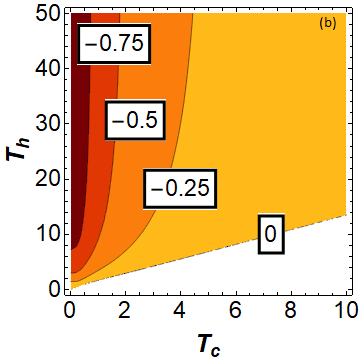}%
}
\subfigure{
     \includegraphics[width=4.1cm,height=4.1cm]{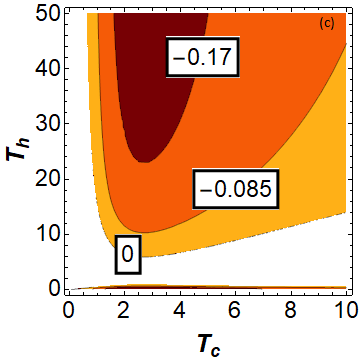}%
}
\subfigure{
     \includegraphics[width=4.1cm,height=4.1cm]{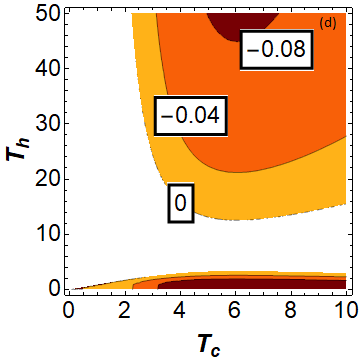}%
}
\caption{Work output $\langle W \rangle$ for a given range of hot and cold temperatures and fixed external magnetic fields $h_i = 3$ and $h_f = 4$. Each figure is plotted for a specific value of coupling strength: (a)~$J=0$; (b)~$J=2$; (c)~$J=5$ and (d)~$J=10$. Darker tones indicate more negative $\langle W \rangle$ (i.e., greater work output). In the white regions the cycle does not operate as a heat engine.}
\label{FigWThTc}
\end{figure}

Let us first discuss the average work $\langle W \rangle$. In Fig.~\ref{FigWThTc} we show a contour plot of $\langle W \rangle$ as a function of $T_h$ and $T_c$ for different values of the coupling $J$. First of all, a clear qualitative distinction can be seen between the top two plots, valid for low $J$, and the bottom two, valid for high $J$. As discussed in detail in ref. \cite{oliveira2021efficiency}, the crossover between the two regimes occurs when $J > h$, and the energy level $E = -J$ becomes the ground state. 

For weak coupling,  there is one continuous parameter region for engine operation, which shrinks as $J$ increases. For strong coupling, two disjoint ``islands" appear, with interesting features. For example; i) in the upper ``island" a decrease in the cold bath temperature $T_c$ can decrease the work output; indeed, engine operation always becomes impossible in the limit $T_c \rightarrow 0$, where in contrast the Carnot efficiency would tend to 1. ii) in the lower ``island" $T_c$ is, in fact, higher than $T_h$. What this means is that the baths exchange their roles: now, after the increase in the external field, we put the system in contact with the colder bath, instead of the hotter one (see  Fig. \ref{Fig:Otto}(b)). In other words, in this temperature range the cycle is `counter-rotating', as described at the beginning of section~\ref {sec:Otto}, and yet still operates as a heat engine. These properties can be understood exclusively in terms of the structure of the energy levels and are not directly related to quantum correlations between the spins \cite{oliveira2021efficiency}. 

Let us now analyze what are regimes of operation that maximize the work output.
Consider first how $\langle W \rangle$ varies with the coupling strength $J$. In Fig.~\ref{FigWThTc}(a) we can see that, for $J=0$, an increase in $T_h$ or a decrease in $T_c$ leads to more negative $\langle W \rangle$ (i.e., greater output), as perhaps would be intuitively expected. Indeed, this behavior can be deduced from eq.(\ref{work}). In particular, the maximum (absolute value) of work that can be obtained is $\Delta h$, in the limits $T_h \rightarrow \infty$ and $T_c \rightarrow 0$. A small value of $J$ (Fig.~\ref{FigWThTc}(b)) does not appear to greatly change this behavior. For most values of $T_h$ and $T_c$ (but not in fact all, see below), $\left. \partial |\langle W \rangle| / \partial J\right|_{T_c, T_h} < 0$ (i.e., increasing $J$ reduces work output). This is due to the partition functions $Z_{c,h}$ appearing in eq.(\ref{work}), both of which increase roughly exponentially with $J$. As a consequence, turning on a small coupling reduces the temperature ranges in which a significant amount of work is produced. (Compare Figs.~\ref{FigWThTc}(a) and ~\ref{FigWThTc}(b)). Nevertheless, the region of maximum work extraction is still the one with large $T_h$ and low $T_c$. 

\begin{figure}[t]
\includegraphics[width=\columnwidth]{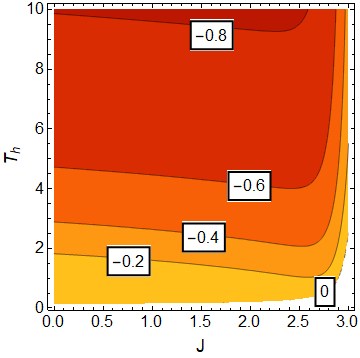}
\caption{Contour plot of $\langle W \rangle$ as a function of $T_h$ and $J$ in the weak-coupling regime, for $h_i = 3$, $h_f = 4$ and $T_c=10^{-1}$}.
\label{Fig-W-Th-J}
\end{figure}

It is however also possible for the output work $|\langle W \rangle|$ to increase with $J$. This happens in particular for small values of $T_h$ and very small values of $T_c$ (i.e., in the left lower corner of Fig.\ref{FigWThTc}(a,b)). This can be seen in Fig. \ref{Fig-W-Th-J}, where we plot contours of $\langle W \rangle$ as a function of $J$ and $T_h$, for a fixed (very small) value of $T_c$. In the region where the contours slope downward, an increase in  $J$ at fixed $T_h$ results in a greater $|\langle W \rangle|$. 

\begin{figure}[h!]
\includegraphics[width=\columnwidth]{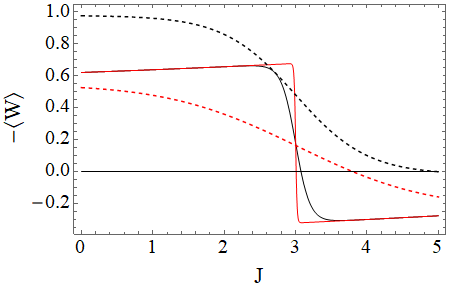}
\caption{Average work output ( $-\langle W \rangle$ ) as a function of the coupling $J$ in the weak-coupling regime. The dashed curves are for ($T_h=100$, $T_c=0.5$) (black, upper) and ($T_h=5$, $T_c=1$) (red, lower). In these cases, the work output decreases monotonically with $J$ (i.e.,  $\langle W \rangle$ becomes monotonically less negative). The solid curves are for $T_c=10^{-1}$ (black, curvier) and $T_c=10^{-2}$ (red, straighter) both for $T_h=5$: in this case, output work can increase with $J$ (i.e., $\langle W \rangle$ can become more negative). Maximum output is reached at a point that approaches the energy level crossing $J = h_i$ as $T_c$ decreases. In all cases, $h_i = 3$, $h_f = 4$. Note that regions with $ \langle W \rangle >0$ do not represent a heat engine.}
\label{Fig-W-J}
\end{figure}

Another perspective of this phenomenon can be seen in Fig.~\ref{Fig-W-J}, where we plot $-\langle W \rangle$ as a function of $J$ for different values of the bath temperatures. The dashed curves, valid for large $T_h$, illustrate the most common behavior, where the work output reduces monotonically ($\langle W \rangle$ becomes less negative), with increasing $J$. The solid curves, valid for small $T_h$, illustrate the surprising situation where the work output can increase ($\langle W \rangle$ becomes more negative) with increasing $J$. Note that, from eq.(\ref{work}), the upper limit for work extraction corresponds to the situation with $x_c \rightarrow \infty$ and $x_h \rightarrow 0$, leading to $\left\langle W \right\rangle \rightarrow -\Delta h$. In comparison, we can see that, in Fig.~\ref{Fig-W-J}, the greatest output is attained at a value of $J$ that converges to the level crossing limit (at $J = h_i = 3$) as $T_c$ decreases. 
Even this maximum remains well under the upper limit, which in this case is $\Delta h = 1$.

\begin{figure}[t]
\includegraphics[width=\columnwidth]{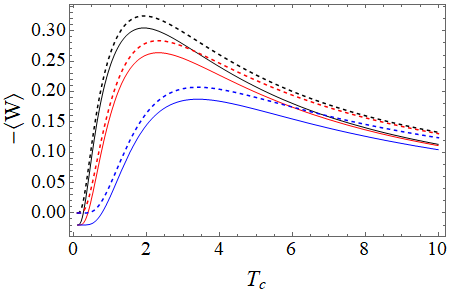}
\caption{Average work output ( $-\langle W \rangle$ ) as a function of $T_c$ in the strong coupling regime, for fixed $T_h$ and different values of $J$.  Solid lines represent a horizontal scan across the `upper island' of plots analogous to those in Fig.~\ref{FigWThTc}(c,d), with $T_h = 100 > T_c$. Dashed lines run across the corresponding `lower island', with $T_h = 10^{-4} < T_c$. In both cases, $J=4.01,\; 4.5$ and $6$ from top curve to bottom. Other parameters are:  $h_i = 3$ and $h_f = 4$.}
\label{Fig:W-T-Islands}
\end{figure}

Turning now to the strong-coupling regime, we find different behaviors, see Fig.\ref{FigWThTc}(c-d). In the upper ``island",  $|\langle W \rangle|$ still decreases with increasing $J$ for most temperature values, and increases in absolute value with the hot bath temperature ($T_h$), as perhaps might be intuitively expected. However, $|\langle W \rangle|$ is not monotonic with respect to the cold bath temperature $T_c$, see the solid lines in Fig.~\ref{Fig:W-T-Islands}. Note that $\langle W \rangle \rightarrow 0$ for both sufficiently low and sufficiently high $T_c$, but reaches a minimum (i.e., the work output $-\langle W \rangle$ reaches a maximum) at some intermediate temperature. The reasons for this non-monotonic behavior are discussed in ref. \cite{oliveira2021efficiency}.

Conversely, in the lower ``island" (counter-rotating engine), $|\langle W \rangle|$ increases as the cold bath temperature ($T_h$, in this case) decreases, as might perhaps be expected.  However, $|\langle W \rangle|$ is not monotonic with respect to the hot bath temperature ($T_c$, in this case), see the dashed lines in Fig.~\ref{Fig:W-T-Islands}. Once again, $\langle W \rangle \rightarrow 0$ for both sufficiently low and sufficiently high $T_c$, but reaches a minimum (i.e., the work output $-\langle W \rangle$ reaches a maximum) at some intermediate temperature. 

Note however that, for both `ìslands'' of the strong-coupling regime, the absolute value of the extracted work is inherently limited. In particular, it can never reach the maximum possible value $\Delta h$. For example, in the conditions of Fig.~\ref{Fig:W-T-Islands}, the greatest value reached by $-\langle W \rangle$ is around $0.34 < \Delta h = 1$. This limitation arises because, as the name implies, the strong-coupling regime only occurs above a minimum threshold for $J$. As already mentioned, this tends to reduce  $|\langle W \rangle|$ 
 due to the exponential increase in the partition functions in eq.~(\ref{work}).

In addition, in this regime, it is normal for the work output to actually increase with increasing $J$. In particular, this is illustrated by the fact that the size of the lower `island', where $\langle W \rangle < 0$, increases from Fig.~\ref{FigWThTc}(c) to Fig.~\ref{FigWThTc}(d). An asymptotic analysis of the boundaries of this region, proving that it does expand with $J$, can also be found in ref. \cite{oliveira2021efficiency}.

\begin{figure}[h!]
\centering
\subfigure{
    \includegraphics[width=0.47\columnwidth]{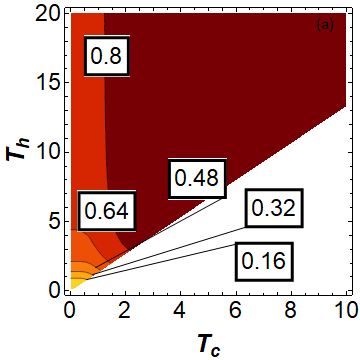}%
}
\subfigure{
\includegraphics[width=0.47\columnwidth]{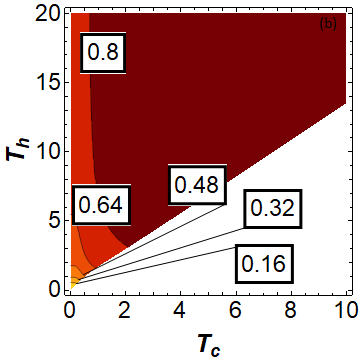}
}
\subfigure{
     \includegraphics[width=0.47\columnwidth]{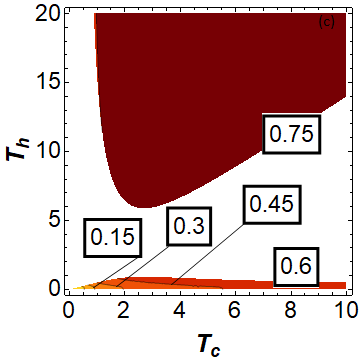}%
}
\subfigure{
     \includegraphics[width=0.47\columnwidth]{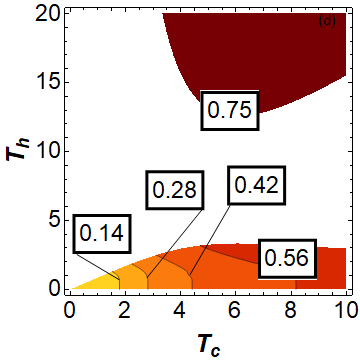}%
}
\caption{Work Fluctuation $\sigma_W$ for a given range of hot and cold temperatures and fixed external magnetic fields $h_i = 3$ and $h_f = 4$. Each figure is plotted for a specific value of coupling strength: (a)~$J=0$; (b)~$J=2$; (c)~$J=5$ and (d)~$J=10$. Darker tones indicate larger $\sigma_W$. In the white regions the cycle does not operate as a heat engine.}
\label{Fig-FluctW}
\end{figure}

\subsection{Work fluctuations}

For macroscopic heat engines, it is natural to identify the output work with its average value, since relative fluctuations become negligible. This is not the case for microscopic engines such as the one being discussed here, where we can expect that fluctuations will be on the order of the average output work itself. In order to actually extract useful work from such an engine, it is paramount therefore to search for parameter regions where the relative fluctuations are as small as possible.  

Let us first consider the absolute fluctuations $\sigma_W$, given by eq.(\ref{eq:fluc}). As discussed in section \ref{sec:mag}, in each stroke $\sigma_{W_i} \rightarrow 0$  for $x_\alpha \rightarrow \infty$. In fact, it can be checked that $\sigma_{W_i} \geq 0$ and decreases monotonically with $x_\alpha$ (i.e, increases monotonically with the temperature $T_\alpha$). In plain words, increasing the temperature of either heat bath always increases the work fluctuations in the cycle.

To illustrate this behavior, in Fig.~\ref{Fig-FluctW} we plot $\sigma_W$ for the same parameter ranges used in Fig.~\ref{FigWThTc}, but using smaller temperature ranges for clarity. As expected, one can see that, both in the weak and strong coupling regimes, $\sigma_W$ is indeed smallest where $T_h$ and $T_c$ are both very small. 

\begin{figure}[h]
\centering
\subfigure{
    \includegraphics[width=0.47\columnwidth]{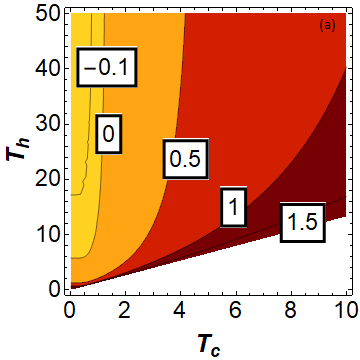}%
}
\subfigure{
\includegraphics[width=0.47\columnwidth]{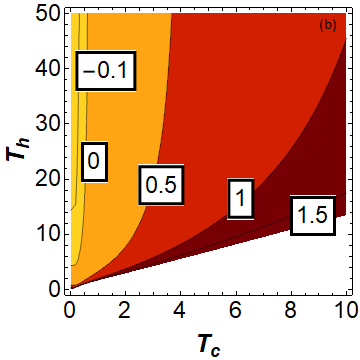}
}
\subfigure{
     \includegraphics[width=0.47\columnwidth]{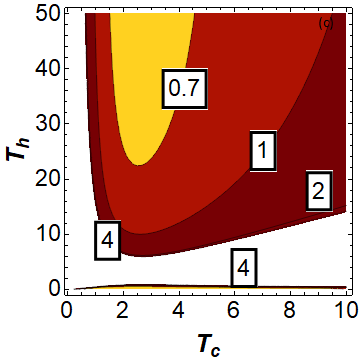}%
}
\subfigure{
     \includegraphics[width=0.47\columnwidth]{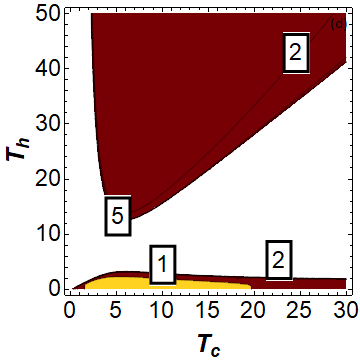}%
}
\caption{Log of the coefficient of variation $(\log |\sigma_{\langle W \rangle}/\langle W \rangle|)$ 
for a given range of hot and cold temperatures and fixed external magnetic fields $h_i = 3$ and $h_f = 4$. Each figure is plotted for a specific value of coupling strength: (a)~$J=0$; (b)~$J=2$; (c)~$J=5$ and (d)~$J=10$. Darker tones indicate larger relative fluctuations. Note that different shading scales have been used in each figure.}
\label{Fig-Fluct-RelatW}
\end{figure}

A more relevant figure of merit than $\sigma_W$ itself is the relative fluctuation of work, or work coefficient of variation:  $|\sigma_{ W}/\langle W \rangle|$. In the following we refer to this coefficient as being low (high) when it is less (greater) than 1, meaning that the work fluctuation surpasses the mean value of the total extracted work.  Fig.~\ref{Fig-FluctW} indicates that one may expect to find a low coefficient of variation is the region where $T_h$ and $T_c$ are both small. However, in this case the work output (the denominator of the coefficient) also becomes very small. In order to avoid numerical instabilities in such situations, it is more convenient to study the logarithm $\log |\sigma_{\langle W \rangle}/\langle W \rangle|$.  In Fig.~\ref{Fig-Fluct-RelatW}, we plot this quantity, using the same temperature ranges, magnetic fields and coupling strengths as in Fig.\ref{FigWThTc}.

In the weak-coupling regime (Fig.~\ref{Fig-Fluct-RelatW}(a,b)), we can see that the cycle operates with low relative work fluctuations (i.e., negative values in this logarithmic plot) only within a very narrow region of high $T_h$ and low $T_c$. Note this happens even for uncoupled spins. Moreover, in the strong-coupling regime we do not find any region with low relative work fluctuations. It seems that, in most cases, the relative fluctuations are large mostly due to the fact that  $\langle W \rangle$ is small, and not because $\sigma_W$ is large.

Let us focus first then on the parameter range with low relative work fluctuations (weak coupling, large $T_h$ and low $T_c$). Note that, in  the limit where both $T_h \rightarrow \infty$ and $T_c \rightarrow 0$ eqs.(\ref{work}) and (\ref{eq:fluc}) give:
\begin{equation}
\begin{split}
    \langle W \rangle &\rightarrow -\Delta h \\
    \sigma_W^2 &\rightarrow (\Delta h)^2/2,
\label{eq:asympwork}
\end{split}
\end{equation}
leading to a coefficient of variation equal to $1/ \sqrt{2}$. Numerical simulations indicate that this is the lowest possible value of this quantity in our system, although we have not been able to prove this analytically. Note that it is independent of the magnetic field and of the interaction coupling. 

\begin{figure}[h]
\includegraphics[width=\columnwidth]{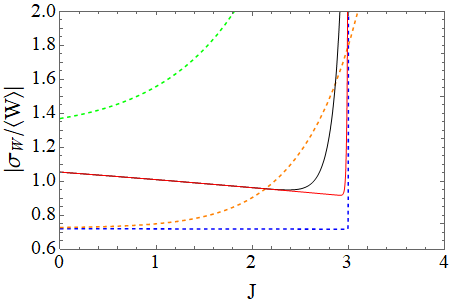}
\caption{Relative fluctuation of work as function of the coupling strength $J$, in the weak-coupling regime, with external magnetic fields $h_i = 3$ and $h_f = 4$. The dashed curves are for $T_h=100$, $T_c=10^{-3}$ (blue, bottom), $T_h=100$, $T_c=0.5$  (orange, middle) and $T_h=5$, $T_c=1$ (green, top).  The solid curves are for $T_h=5$, $T_c=10^{-1}$ (black, top) and $T_c=10^{-2}$ (red, bottom). In this case the relative fluctuations can decrease as $J$ increases, reaching a minimum at a point that approaches the energy level crossing as $T_c$ decreases.}
\label{Fig-FlucW-J-LowTh}
\end{figure}

Let us now discuss how the spin-spin coupling affects the relative fluctuations of work in the weak-coupling regime. Investigating numerically, we find the following (Fig.~\ref{Fig-FlucW-J-LowTh}). The most common behavior, valid for most bath temperatures in the ranges represented in Fig.~\ref{Fig-Fluct-RelatW}, is very similar to that of the average work,  i.e., the relative fluctuation increases with $J$. This is represented by the dashed curves in Fig.~\ref{Fig-FlucW-J-LowTh}. In the upper dashed curve (green) $T_h=5$ and $T_c=1$, the middle dashed curve (orange) is for $T_h=100$ and $T_c=0.5$, and the bottom
dashed (blue) curve is for $T_h=100$ and $T_c=10^{-3}$. We can see that, for large  $T_h$ and small $T_c$, the relative fluctuations approach the minimum value $1/\sqrt{2}$, staying close to it for a range of $J$ that tends to the level crossing point, $J = h_i = 3$ as $T_c$ decreases. In particular, for very small $T_c$ (blue curve), the fluctuations stay practically at the minimum almost right up to the level crossing point.

There are also cases, however, where the relative fluctuation can actually decrease with $J$, although by a small amount. This can be seen in the two solid curves, for $T_h=5$ with $T_c=10^{-1}$ (black, upper) and $T_c=10^{-2}$ (red, lower). In these cases the relative fluctuations reach a minimum at a point that
approaches the energy level crossing $J = h_i = 3$ as $T_c$ decreases, and that can go under 1.

Finally, in the strong coupling limit the relative fluctuation is always larger than 1. In both "islands" of Fig.\ref{Fig-Fluct-RelatW}(c-d),it displays the same qualitative behavior with the bath temperatures as the average work $\langle W \rangle$. Note that, even in the lower ``islands''  where the absolute fluctuation can be very small (see Fig.~\ref{Fig-FluctW}), the relative fluctuation is nevertheless always above 1.

\section{Efficiency}
\label{eff}

Up to this point we have focused on identifying the regions in parameter space where the Otto cycle outputs large work with small relative fluctuations. We have found this occurs most favorably for weak antiferromagnetic coupling, low $T_c$ and high $T_h$, although there are also other regimes. In this section, we explore how efficient this process is, focusing on these same regions of parameter space. In particular, here we assume $T_h > T_c$ and $0\leq J < h_i < h_f$.

\subsection{Efficiency fluctuation}

As discussed in the introduction, we will consider efficiency as a stochastic quantity, using the notion of `scaled fluctuating efficiency' $\eta_{s'}$ in eq.(\ref{stochasticefficiency2}) \cite{fei2022efficiency}.
As shown in e.g. \cite{thomas2011coupled,oliveira2021efficiency}, the average heat absorbed from the hot bath in this cycle may be written as
\begin{equation}
\langle Q_h \rangle = J (p_c^J - p_h^J) - \frac{h_f}{\Delta h} \langle W \rangle.
\end{equation}
It can then be checked \cite{thomas2011coupled,oliveira2021efficiency,anka2021measurement} that the average value of $\eta_{s'}$, corresponding to the usual thermodynamic efficiency (eq.(\ref{themodeff})), is

\begin{equation}
\eta_{th} = \dfrac{\eta_0}{1 + (J/h_f) \Omega},
\label{effth}
\end{equation}
where $\eta_0 = 1 - h_i/h_f$ is the standard quantum Otto efficiency for a WS composed of a single spin in a magnetic field. Here 
\begin{equation}
\Omega \equiv \dfrac{p_h^J - p_c^J}{\langle M_h^f \rangle - \langle M_c^i \rangle}
\end{equation}
is a quantity that can have either sign. Thus, it is possible for Otto cycles in this system to operate with an efficiency either higher or lower than $\eta_0$  \cite{thomas2011coupled,oliveira2021efficiency}. For $J = 0$ (non-interacting spins) we recover $\eta_{th} = \eta_0$ as expected.  As discussed in \cite{oliveira2021efficiency}, the situations where $\eta_{th} > \eta_0$ can be interpreted as being due to an increase in the heat flowing from the hot bath to the cold one via the `working' levels, at the expense of heat flowing `in reverse' (from cold to hot) via the `idle' levels. It can be checked explicitly that this expression for $\eta_{th}$  is indeed always smaller than the Carnot efficiency corresponding to the two bath temperatures, as it should \cite{thomas2011coupled}.

 Let us now consider the fluctuations in the thermodynamic efficiency. Using eqs.(\ref{varcoefficient}) and (\ref{workfluct}), we can write

\begin{equation}
\begin{split}
\sigma^2_{\eta_{s'}} & = \eta_{th}^2 \dfrac{\sigma^2_{ M_c^i} + \sigma^2_{ M_h^f }}{\Delta M^2} \\
 & = \eta_0^2 \left[ \dfrac{\sigma^2_{ M_c^i } + \sigma^2_{ M_h^f }}{\Delta M^2\Big(1 + (J/h_f)\Omega\Big)^2} \right]
\end{split}
\end{equation}

In particular, in the limit where $T_c \rightarrow 0$ and $T_h \rightarrow \infty$ we have:
\begin{align}
 p_c^J &\rightarrow 0; \;\;\;\;\; p_h^J  \rightarrow 1/4 \nonumber \\ 
 \langle M_c^i \rangle &\rightarrow 1;\;
 \langle M_h^f \rangle \rightarrow 0\\
 \sigma^2_{M_c^i} &\rightarrow 0;\;
\sigma^2_{M_h^f} \rightarrow 1/2 \nonumber
 \end{align}
and so
\begin{equation}
\eta_{th} \rightarrow \dfrac{\eta_0}{1 - J/4h_f},  \\
\label{efflim}
\end{equation}
\begin{equation}
\sigma_{\eta_{s'}}^2 \rightarrow \dfrac{\eta_0^2}{2(1 - J/4h_f)^2}.
\end{equation}

Note that, in contrast with the corresponding expressions for $\left\langle W\right\rangle$ and $\sigma_W$ in eq.(\ref{eq:asympwork}), the asymptotic values of the efficiency and its fluctuations still depend on the coupling strength $J$. Moreover,  since $J\geq 0$, we can see that in this limit $\eta_0~\leq~\eta_{th}$, i.e., the spin-spin interaction always has a positive impact on the cycle performance \footnote{For a more detailed analysis of this asymptotic behavior, taking into account finite but nonzero values of $T_c$, see Appendix B, section 2 of \cite{oliveira2021efficiency}.}. Finally, since $J < h_i$, then of course $\eta_{th} < 1$, which is the Carnot efficiency in this limit. 

\begin{figure}[h!]
\centering
\subfigure{
    \includegraphics[width=0.47\columnwidth]{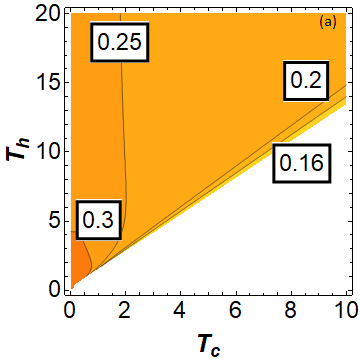}%
}
\subfigure{
\includegraphics[width=0.47\columnwidth]{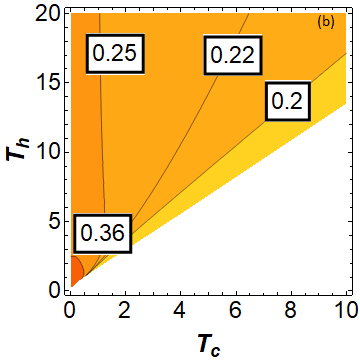}
}
\subfigure{
     \includegraphics[width=0.47\columnwidth]{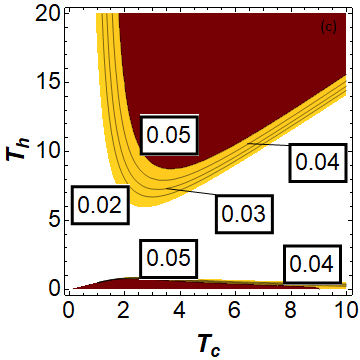}%
}
\subfigure{
     \includegraphics[width=0.47\columnwidth]{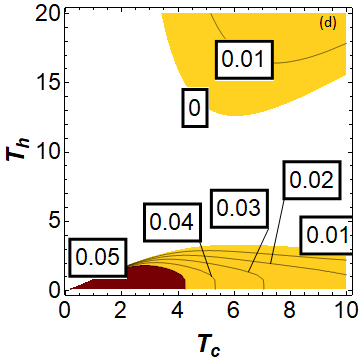}%
}
\caption{Thermodynamic efficiency $\eta_{th}$ for a given range of hot and cold temperatures and fixed external magnetic fields $h_i = 3$ and $h_f = 4$. Each figure is plotted for a specific value of coupling interaction: (a) $J=1.5$, (b) $J=2$, (c) $J=5$ and (d) $J=10$.}
\label{Fig-Eff}
\end{figure}

In Fig.~\ref{Fig-Eff} we plot the thermodynamic efficiency $\eta_{th}$  within a range of hot and cold bath temperatures, for various coupling strengths: (a) $J=1.5$, (b) $J=2$, (c) $J=5$ and (d) $J=10$. In all cases, the extremal magnetic fields are $h_i=3$ and $h_f=4$. Note first that $\eta_{th}$ may reach values that are both higher and lower than the standard Otto value $\eta_0$ (here $=0.25$). Note also that, for a given value of $J$ in the weak-coupling regime ($J < h_i = 3$), the efficiency can actually increase as the temperature gap between the baths decreases, a counter-intuitive property discussed in Ref.\cite{oliveira2021efficiency}.
In the strong-coupling regime ($J>h_f = 4)$ we can see that $\eta_{th}$ 
is nearly zero for the whole range of temperatures. This is due to the small amount of work extracted by the counter-rotating engine, as shown in Fig.\ref{FigWThTc}(c-d).

As mentioned in section \ref{intro}, the fluctuations of the scaled fluctuating efficiency, as defined in eq.(\ref{stochasticefficiency2}), are inherited from those of work, leading to eq.(\ref{varcoefficient}). Thus, the region of lower relative fluctuations for the thermodynamic efficiency is exactly the same as that shown in Fig.~\ref{Fig-Fluct-RelatW}. Indeed, comparing Fig. \ref{Fig-Fluct-RelatW} with Fig.~\ref{Fig-Eff}, we see that, in this same region, it is possible to achieve low relative work fluctuations while still maintaining the efficiency higher than the standard Otto value $\eta_0$. However, it must be emphasised that these cycles produce no power, since we consider perfect thermalization ($\tau \rightarrow \infty$).

\begin{figure}[h]
\centering
\subfigure{
    \includegraphics[width=1\columnwidth]{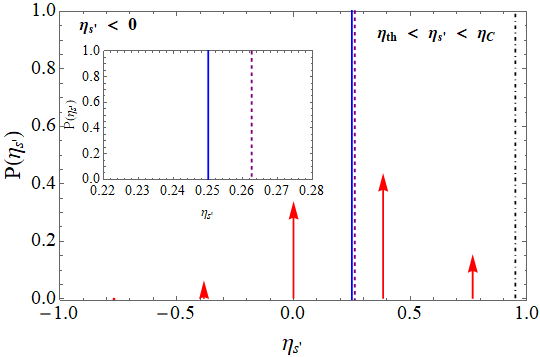}%
}
\subfigure{
\includegraphics[width=1\columnwidth]{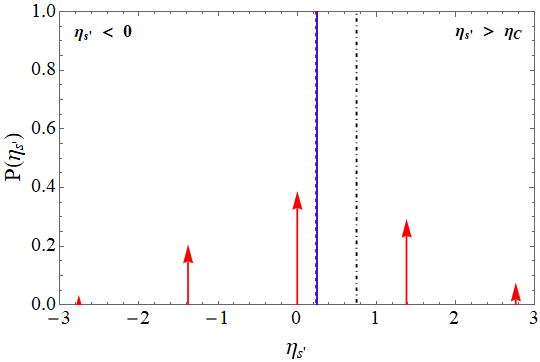}
}
\caption{Probability distribution of the thermodynamic efficiency for a set of parameters: $h_i=3$, $h_f=4$, $T_h=20$, $J=1.5$, $T_c=1$ (top) and $T_c=5$ (bottom). The blue (solid), purple (dashed), and black (dot-dashed) vertical lines represent the Otto, thermodynamic, and Carnot efficiencies, respectively. Note that, for the parameters of the top figure, the thermodynamic efficiency is only very slightly higher than the Otto value. A magnified view of the difference between them is shown in the inset. In the bottom figure, the two are essentially indistinguishable.}
\label{effprob}
\end{figure}

In order to illustrate the influence of the stochastic nature of the TPM protocol on the efficiency fluctuations of the cycle, we show in Fig.~\ref{effprob} two probability distributions, eq.(\ref{stochasticefficiency}), for the thermodynamic efficiency, eq.(\ref{stochasticefficiency2}), at different cold bath temperatures: $T_c=1$ (top) and $T_c=5$ (bottom). All other parameters are fixed:  $h_i=3$, $h_f=4$, $T_h=20$, $J=1.5$. 
In either case, the blue vertical line represents the standard Otto efficiency, the grey line  the Carnot efficiency, and the purple line the actual average thermodynamic efficiency. It is clear that, in both cases, the stochastic values of the thermodynamic efficiency fluctuate quite far from their average. In fact it is possible to obtain positive, negative or zero values of $\eta_{s'}$, with finite probability. Null efficiency simply represents the case where no net work is extracted. Negative efficiency means that, in a particular run, the cycle can actually require input work from the external work `sink' in order to run. Of course, since the average thermodynamic efficiency is positive in an engine cycle, the distribution must skew towards the positive stochastic values of $\eta_{s'}$. Note however that the width of the distribution depends on $T_c$. For low $T_c$ (top) the stochastic efficiency may only achieve positive values that are in between the thermodynamic and Carnot efficiency. For high enough $T_c$ (bottom), however, they can be even higher than the Carnot efficiency. This is in agreement with the discussion of the previous sections, where it was shown that an increase in $T_c$ leads to a higher fluctuation.

\subsection{Entropic bound}
\begin{figure}[ht!]
\includegraphics[width=\columnwidth]{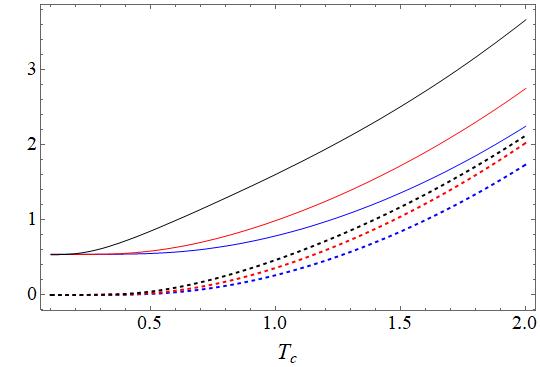}
\caption{Comparison between the left and right sides of eq.(\ref{eqtur}): solid lines show the exact relative work (or efficiency) fluctuation $\sigma^2_{\eta'}/\eta_{th}^2 =\sigma^2_W/\langle W \rangle^2$, and dotted lines the TUR-derived lower bound $f(\langle \Sigma \rangle)$. In both cases, $J$ decreases from top to bottom: $J=2$ (black, top), $J=1$ (red, middle) and $J=0$ (blue, bottom). In all cases, $h_i=3$, $h_f=4$ and $T_h=50$.}
\label{Fig-TUR}
\end{figure}

General bounds on work and efficiency fluctuations (or those of any other thermodynamic current) can be obtained from so-called Thermodynamic Uncertainty Relations (TURs), which link these quantities to the average entropy production $\langle \Sigma \rangle$ \cite{barato2015thermodynamic,timpanaro2019thermodynamic,proesmans2017discrete,horowitz2020thermodynamic}.  Although many different bounds of this kind can be constructed \cite{horowitz2020thermodynamic}, here we use the one derived by Timpanaro \textit{et. al.}, which is the tightest possible general bound \cite{timpanaro2019thermodynamic}. It is given by
\begin{equation}
\begin{split}
\dfrac{\sigma^2_{\eta'}}{\eta_{th}^2} &=\dfrac{\sigma^2_W}{\langle W \rangle^2} \geq f(\langle \Sigma \rangle),
\end{split}
\label{eqtur}
\end{equation}
where $f(x) = \csch^2[g(x/2)]$, $\csch(x)$ is the hyperbolic cosecant, and $g(x)$ is the inverse function of $x \tanh(x)$. 
In Fig.\ref{Fig-TUR} we illustrate this bound, showing that it is indeed respected. 
Note however that it is not saturated in the limit $T_c \rightarrow 0$. This happens since, while on one hand $\langle \Sigma \rangle \rightarrow \infty$ in this limit, and hence the bound $f(\langle \Sigma \rangle) \rightarrow 0$, on the other the fluctuation in the total work (solid lines) remains finite, due to the contribution arising from the stroke starting from equilibrium with the hot bath. Note also that the apparent convergence of these fluctuations to a single value at $T_c \rightarrow 0$ is merely an artifact of the scale of the graph; the limiting values for different $J$ remain different. 

\begin{figure}[th!]
\centering
\subfigure{
    \includegraphics[width=0.45\columnwidth]{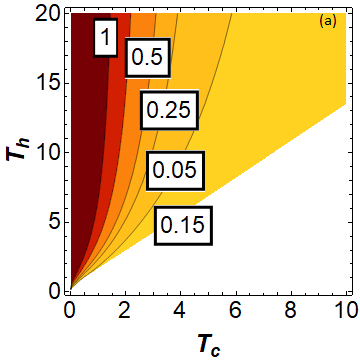}
}
\subfigure{
\includegraphics[width=0.45\columnwidth]{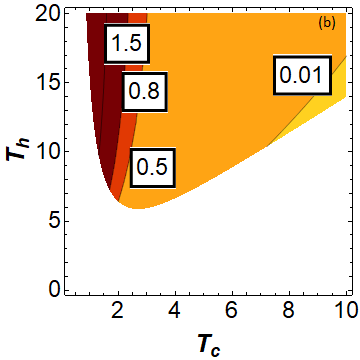}
}
\caption{Entropy production $\langle \Sigma \rangle$ for $J=2$ (a) and $J=5$ (b).}
\label{tur}
\end{figure}

More broadly speaking, in the case of the Otto cycle, entropy is only produced in the isochoric strokes, and so it is easy to show that 
\begin{equation}
\langle \Sigma \rangle = \beta_c \langle Q_h \rangle (\eta_C - \eta_{th}),
\end{equation}
where $\eta_h$ is given in eq.(\ref{effth}). 
It follows then from eq.(\ref{eqtur}) that, whenever the thermodynamic efficiency approaches the Carnot efficiency, relative work fluctuations must become large. Given eq.(\ref{varcoefficient}), the same is also true of the relative fluctuations in the thermodynamic efficiency. This is illustrated in Fig.~\ref{tur}, where we show the entropy production $\langle \Sigma \rangle$ for $J=2$ (ordinary engine regime) and $J=5$ (counter-rotating engine regime). In  Fig.~\ref{tur}(a), it is possible to see that $\langle \Sigma \rangle$ is larger in the same region of the diagram (high $T_h$ and low $T_c$) where we find lower relative fluctuations in work (compare Fig.~\ref{Fig-FluctW}(b)). In other words, the more irreversible the cycle is, the less relative work fluctuation is observed. Meanwhile, in the counter-rotating regime (Fig.~\ref{tur}(b))  the work extraction is so small that the relative fluctuations still remain substantial even in the region of large entropy production (compare Fig.~\ref{Fig-FluctW}(c)).

\section{Conclusion}
\label{conclusion}

In this work, we have studied the performance of a quantum Otto engine with two interacting qubits as its working substance, taking into account fluctuations. We first proved relations between the mean work and the mean magnetization at equilibrium and between the fluctuation of work and the magnetic susceptibility at equilibrium. Although such relationships are expected for thermodynamical systems at equilibrium, it is interesting that they remain true in our case, where the working system is driven out of equilibrium after each unitary stroke. 

We showed that the maximum average work output for this engine is achieved in the `regular engine'  regime (small $J$), with high hot bath temperature and low cold bath temperature, as one would expect for classical engines. In general, we found that, except for a small parameter region where both bath temperatures are small, the mean output work always decreases with the size of the coupling constant $J$.
With regard to the work fluctuation, it is the regime with the two baths at low temperatures that presents the smallest values. However, the relative work fluctuation is very large for most bath temperatures, being smaller than 1 only at very high
hot bath temperatures and low cold bath temperatures. We also showed that, similarly to the mean work, the work fluctuations can also sometimes decrease as the coupling constant increases. However, it appears that the relative work fluctuation in this system never dips below $1/\sqrt{2} \simeq 0.7$. Therefore this microscopic engine with only two particles can never really be a reliable work source - as should perhaps be expected. 

Furthermore, we studied the average efficiency $\eta_{th}$ of this cycle in all possible regimes of operation as an engine. We showed that it is very small for large $J$, but can be larger than the standard Otto efficiency for small $J$ and small cold bath temperature. It can even increase with a decrease in the hot bath temperature. 

In particular, we were able to identify a regime where the engine operates both with high performance (large mean work output) and high reliability (low relative fluctuations), while still remaining with efficiency above the standard Otto value. This regime occurs in
the limit of small $J$, large hot bath temperature, and small cold bath temperature.

Finally, we analyzed the probability distribution of the ``scaled fluctuating efficiency" $\eta_{s'}$  \cite{fei2022efficiency} for this cycle, showing that it can achieve values that are greater than both the Otto efficiency and the Carnot efficiency. There is also a nonzero probability of zero efficiency, where no work is extracted, or of negative efficiency, in which case that the total work is positive, i.e., the system is not operating in an engine mode. Nevertheless, the average $\eta_th$ remains below the Carnot bound, as it must.  We also checked that the relative efficiency fluctuations in our cycle respect the lower bound set by the tightest general thermodynamic uncertainty relation derived so far.

\begin{acknowledgments}

This study was financed in part by Coordenação de Aperfeiçoamento de Pessoal de Nível Superior - Brazil (CAPES) - Finance Code 001, Instituto Nacional de Ciência e Tecnologia de Informação Quântica (465469/2014-0), the Conselho Nacional de Desenvolvimento Científico e Tecnológico (CNPq), and the Fundação Carlos Chagas de Amparo à pesquisa do Estado do Rio de Janeiro (FAPERJ). T.R.O. acknowledges the financial support of the Air Force Office of Scientific Research under Award No. FA9550-23-1-0092.

\end{acknowledgments}

\appendix

\bibliography{Workfluctuation}

\end{document}